\documentclass[a4paper,10pt,twoside]{cpc-hepnp}
\usepackage{CJK,upgreek,fancyhdr}
\usepackage{subfigure}
\usepackage{multicol}
\usepackage{booktabs}
\usepackage{amssymb,bm,mathrsfs,bbm,amscd}
\usepackage[tbtags]{amsmath}
\usepackage{lastpage}
\usepackage{graphicx}

\DeclareGraphicsExtensions{.pdf,.jpeg,.png}

\usepackage{epstopdf}

\usepackage[english]{babel}
\usepackage{overpic}
\usepackage{color}

\newcommand{\ee}{e^+e^-}

\newcommand{\br}[1]{\mathcal{B}(#1)}

\begin{document}
\begin{CJK*}{UTF8}{gkai}

\fancyhead[c]{\small Chinese Physics C~~~Vol. xx, No. x (201x) xxxxxx}
\fancyfoot[C]{\small 010201-\thepage}

\footnotetext[0]{Received xxxx June xxxx}

\title{Evidence for the decays of $\Lambda^+_{c}\to\Sigma^+\eta$ and $\Sigma^+\eta^\prime$
\thanks{
Supported in part by National Key Basic Research Program of China under Contract No. 2015CB856700; National Natural Science Foundation of China (NSFC) under Contracts Nos. 11235011, 11275266, 11335008, 11425524, 11625523, 11635010; the Chinese Academy of Sciences (CAS) Large-Scale Scientific Facility Program; the CAS Center for Excellence in Particle Physics (CCEPP); Joint Large-Scale Scientific Facility Funds of the NSFC and CAS under Contracts Nos. U1332201, U1532257, U1532258; CAS under Contracts Nos. KJCX2-YW-N29, KJCX2-YW-N45, QYZDJ-SSW-SLH003; 100 Talents Program of CAS; National 1000 Talents Program of China; INPAC and Shanghai Key Laboratory for Particle Physics and Cosmology; German Research Foundation DFG under Contracts Nos. Collaborative Research Center CRC 1044, FOR 2359; Istituto Nazionale di Fisica Nucleare, Italy; Koninklijke Nederlandse Akademie van Wetenschappen (KNAW) under Contract No. 530-4CDP03; Ministry of Development of Turkey under Contract No. DPT2006K-120470; National Science and Technology fund; The Swedish Research Council; U. S. Department of Energy under Contracts Nos. DE-FG02-05ER41374, DE-SC-0010118, DE-SC-0010504, DE-SC-0012069; University of Groningen (RuG) and the Helmholtzzentrum fuer Schwerionenforschung GmbH (GSI), Darmstadt; WCU Program of National Research Foundation of Korea under Contract No. R32-2008-000-10155-0
}}

\maketitle
\begin{center} 
\begin{small}
\begin{center}
M.~Ablikim(麦迪娜)$^{1}$, M.~N.~Achasov$^{10,d}$, S. ~Ahmed$^{15}$, M.~Albrecht$^{4}$, M.~Alekseev$^{56A,56C}$, A.~Amoroso$^{56A,56C}$, F.~F.~An(安芬芬)$^{1}$, Q.~An(安琪)$^{43,53}$, Y.~Bai(白羽)$^{42}$, O.~Bakina$^{27}$
% ...

\vspace{0.2cm}
(BESIII Collaboration)\\
\vspace{0.2cm} {\it
$^{1}$ Institute of High Energy Physics, Beijing 100049, People's Republic of China\\
% ...
$^{4}$ Bochum Ruhr-University, D-44780 Bochum, Germany\\
% ...
$^{10}$ G.I. Budker Institute of Nuclear Physics SB RAS (BINP), Novosibirsk 630090, Russia\\
% ...
$^{15}$ Helmholtz Institute Mainz, Johann-Joachim-Becher-Weg 45, D-55099 Mainz, Germany\\
% ...
$^{27}$ Joint Institute for Nuclear Research, 141980 Dubna, Moscow region, Russia\\
% ...
$^{42}$ Southeast University, Nanjing 211100, People's Republic of China\\
$^{43}$ State Key Laboratory of Particle Detection and Electronics, Beijing 100049, Hefei 230026, People's Republic of China\\
% ...
$^{53}$ University of Science and Technology of China, Hefei 230026, People's Republic of China\\
% ...
$^{56}$ (A)University of Turin, I-10125, Turin, Italy; (B)University of Eastern Piedmont, I-15121, Alessandria, Italy; (C)INFN, I-10125, Turin, Italy\\
% ...
$^{d}$ Also at the Novosibirsk State University, Novosibirsk, 630090, Russia\\
% ...
}\end{center}

\vspace{0.4cm}
\end{small}
\end{center}


\begin{abstract}
We study the hadronic decays of $\Lambda_{c}^{+}$ to the final states $\Sigma^{+}\eta$ and $\Sigma^+\eta^\prime$, using an $\ee$ annihilation data sample of 567 pb$^{-1}$ taken at a center-of-mass energy of 4.6 GeV with the BESIII detector at the BEPCII collider.
We find evidence for the decays $\Lambda_{c}^{+}\rightarrow\Sigma^{+}\eta$ and $\Sigma^+\eta^\prime$ with statistical significance of $2.5\sigma$ and $3.2\sigma$, respectively.  Normalizing to the reference decays $\Lambda_c^+\to\Sigma^+\pi^0$ and $\Sigma^+\omega$, we obtain the ratios of the branching fractions $\frac{{\mathcal B}(\Lambda_c^+\to\Sigma^+\eta)}{{\mathcal B}(\Lambda_c^+\to\Sigma^+\pi^0)}$ and $\frac{{\mathcal B}(\Lambda_c^+\to\Sigma^+\eta^\prime)}{{\mathcal B}(\Lambda_c^+\to\Sigma^+\omega)}$ to be $0.35 \pm 0.16 \pm 0.03$ and $0.86 \pm 0.34 \pm 0.07$, respectively. The upper limits at the 90\% confidence level are set to be  $\frac{{\mathcal B}(\Lambda_c^+\to\Sigma^+\eta)}{{\mathcal B}(\Lambda_c^+\to\Sigma^+\pi^0)}<0.58$ and $\frac{{\mathcal B}(\Lambda_c^+\to\Sigma^+\eta^\prime)}{{\mathcal B}(\Lambda_c^+\to\Sigma^+\omega)}<1.2$. 
Using BESIII measurements of the branching fractions of the reference decays, we determine $\br{\Lambda_{c}^{+}\rightarrow\Sigma^{+}\eta}=(0.41\pm0.19\pm0.05)\%$ ($<0.68\%$) and $\br{\Lambda_{c}^{+}\rightarrow\Sigma^{+}\eta'}=(1.34\pm0.53\pm0.21)\%$ ($<1.9\%$). Here, the first uncertainties are statistical and the second systematic. 
The obtained branching fraction of $\Lambda_c^+\to\Sigma^+\eta$ is consistent with the previous measurement, and the branching fraction of $\Lambda_{c}^{+}\rightarrow\Sigma^{+}\eta^{\prime}$ is measured for the first time.
\end{abstract}

\begin{keyword}
charmed baryon, $\Lambda_c^+$ decays, branching fractions
\end{keyword}

\begin{pacs}
13.66.Bc, 14.20.Lq, 13.30.Eg
\end{pacs}

\newpage
\begin{multicols}{2}

\section{Introduction}
 Nonleptonic decays of charmed baryons offer excellent opportunities for testing different theoretical approaches to describe the complicated dynamics of heavy-light baryons, including the current algebra approach~\cite{uppal}, the factorization scheme, the pole model technique~\cite{xuqp,sharma,driver}, the relativistic quark model~\cite{korner,ivan} and the quark-diagram scheme~\cite{chenghy2}. Contrary to the significant progress made in the studies of heavy meson decays, the progress in 
 both theoretical and experimental studies of heavy baryon decays is relatively sparse. The $\Lambda_{c}^{+}$ was first observed at the Mark II experiment in 1979~\cite{mark2}, but only about 60\% of its decays have been accounted for so far and the rest still remain unknown~\cite{pdg2016}.

 The two-body Cabibbo-favored (CF) decay of the $\Lambda_{c}^{+}$ to an octet baryon and a pseudoscalar meson, $\Lambda_{c}^{+}\to B(\frac{1}{2}^+)P$, is one of the simplest hadronic channels to be treated theoretically~\cite{chenghy}, and measurements of the branching fractions (BFs) can be used to calibrate different theoretical approaches. 
Recently, BESIII has studied twelve CF $\Lambda_{c}^{+}$ decay modes, among which  the absolute BFs for $B(\frac{1}{2}^+)P$ decays $\Lambda_{c}^{+}\rightarrow p K_{S}^0$, $\Lambda \pi^+$, $\Sigma^0 \pi^+$ and $\Sigma^+ \pi^0$ are significantly improved in precision~\cite{lipr}. However, other CF modes are only known with poor precision, or even have not been explored yet. 
 
  \begin{center}
\begin{minipage}{0.5\textwidth}
\centering
\includegraphics[width=0.48\textwidth]{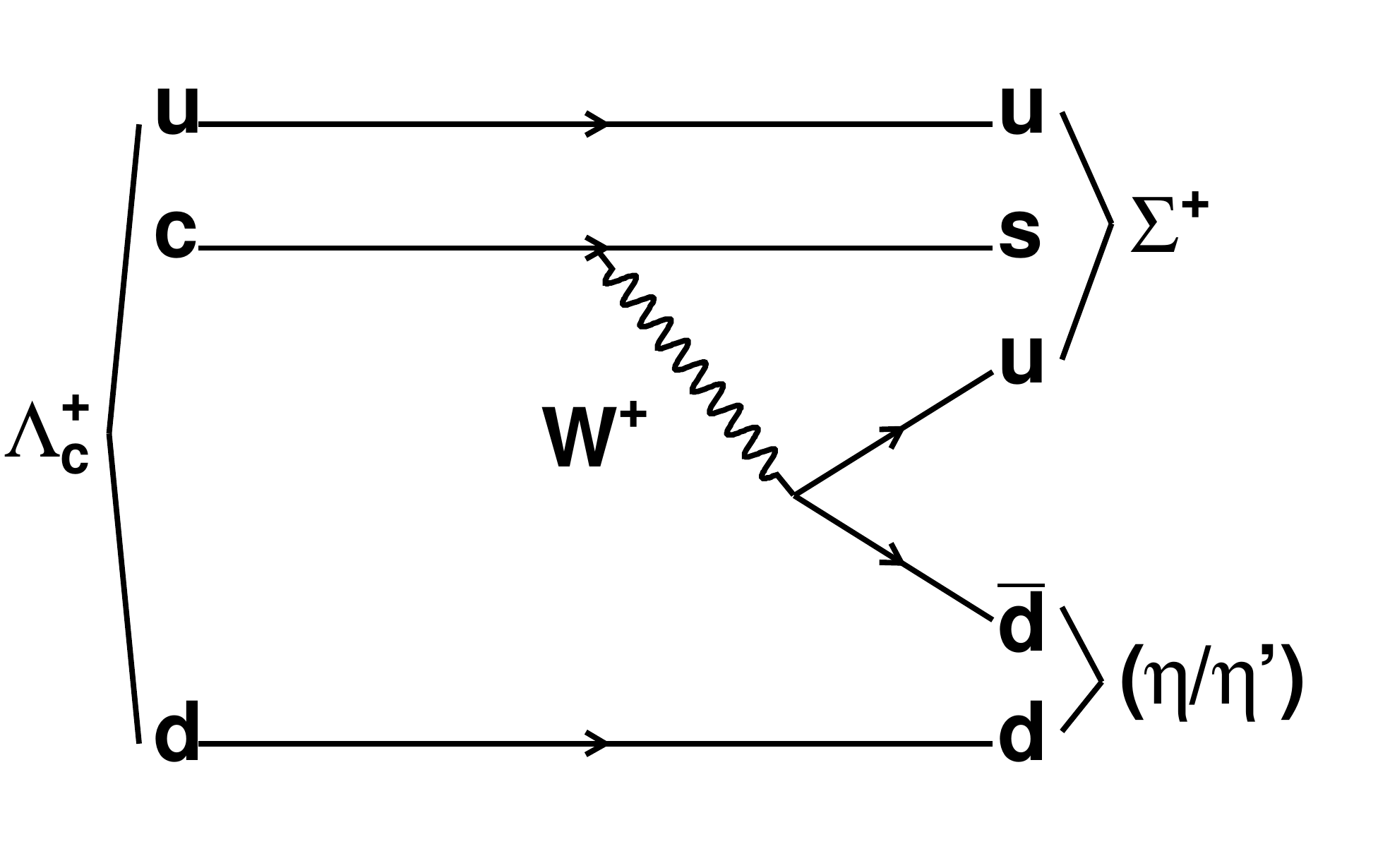}
\includegraphics[width=0.48\textwidth]{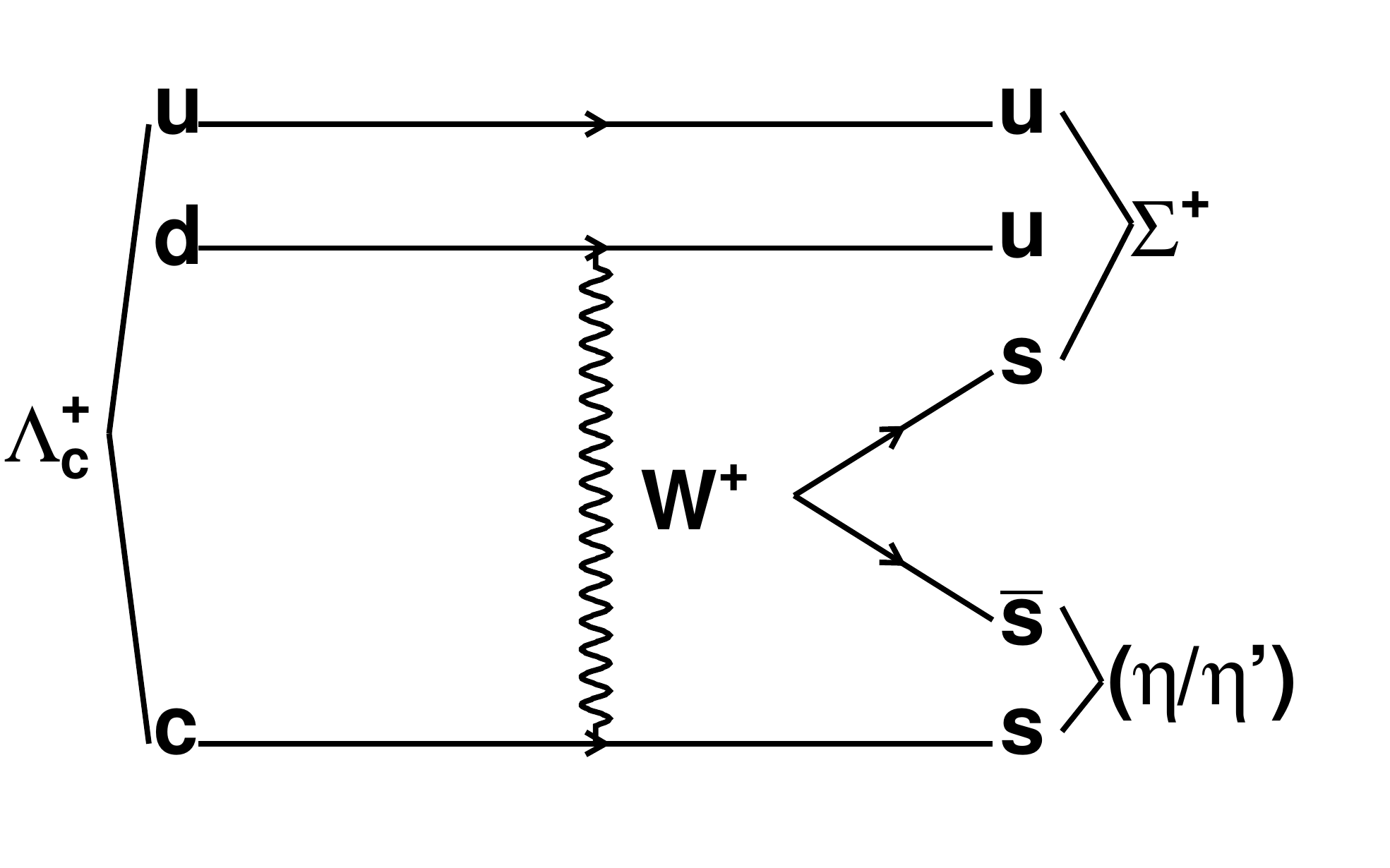}
\figcaption{\label{fig:feynman}Representative tree level diagrams of decays of $\Lambda_{c}^{+}\rightarrow\Sigma^{+}\eta$ and $\Lambda_{c}^{+}\rightarrow\Sigma^{+}\eta^{\prime}$.  }
 \end{minipage}
 \end{center} 
 
 The CF decays $\Lambda_{c}^{+}\rightarrow\Sigma^{+}\eta$ and $\Sigma^{+}\eta^{\prime}$ proceed entirely through nonfactorizable internal $W$-emission and $W$-exchange diagrams, as shown in Fig.~\ref{fig:feynman}, and are particularly interesting. Unlike the case for charmed meson decays, these nonfactorizable decays are free from color and helicity suppressions and are quite sizable. 
Theoretical predictions on these nonfactorizable effects are not reliable, however, resulting in very large variations of the predicted BFs, \emph{e.g.}, ${\mathcal B}(\Lambda_{c}^{+}\rightarrow\Sigma^{+}\eta)=(0.11-0.94)\%$, and ${\mathcal B}(\Lambda_{c}^{+}\rightarrow\Sigma^{+}\eta^{\prime})=(0.1-1.28)\%$~\cite{korner,sharma,driver,ivan}. 
On the experimental side, only evidence for $\Lambda_{c}^{+}\rightarrow\Sigma^{+}\eta$ has been reported by CLEO~\cite{Ammar:1995je} with a BF of $(0.70\pm0.23)$\%, and the channel $\Lambda_{c}^{+}\rightarrow\Sigma^{+}\eta'$ is yet to be observed.
Hence, further experimental studies of these two decay modes are essential for testing different theoretical models and for a better understanding of the $\Lambda_{c}^{+}$ CF decays.

In this work, BFs for $\Lambda_{c}^{+}\rightarrow\Sigma^{+}\eta$ and $\Sigma^{+}\eta^{\prime}$ are measured with respect to the CF modes $\Lambda_{c}^{+}\rightarrow\Sigma^{+}\pi^{0}$ and $\Sigma^{+}\omega$, respectively, by analyzing 567~pb$^{-1}$~\cite{lum} data taken at $\sqrt{s}=4.6$ GeV~\cite{Ablikim:2015zaa} with the BESIII detector at the BEPCII collider. Throughout this paper, charge-conjugate modes are always implied.

\section{BESIII detector}

The BESIII detector has a geometrical acceptance of 93\% of
4$\pi$ and consists of the following main components: 1) a
small-celled, helium-based main draft chamber (MDC) with 43 layers.
The average single wire resolution is 135 $\rm \mu m$, and the
momentum resolution for 1~GeV/$c$ charged particles in a 1 T magnetic
field is 0.5\%; 2) a Time-Of-Flight system (TOF)
for particle identification composed of a barrel part made of two
layers with 88 pieces of 5 cm thick, 2.4 m long plastic scintillator
in each layer, and two end-caps each with 96 fan-shaped, 5 cm thick, plastic
scintillators. The time resolution is 80~ps in the
barrel, and 110 ps in the endcaps, corresponding to a $2\sigma$
K/$\pi$ separation for momenta up to about 1.0 GeV/$c$; 
3) an electromagnetic calorimeter (EMC) made of 6240
CsI~(Tl) crystals arranged in a cylindrical shape (barrel) plus two
end-caps. For 1.0 GeV photons, the energy resolution is 2.5\% in the
barrel and 5\% in the end-caps, and the position resolution is 6 mm in
the barrel and 9 mm in the end-caps;  4) a muon chamber
system (MUC) made of Resistive Plate Chambers (RPC)
arranged in 9 layers in the barrel and 8 layers in the endcaps and
incorporated in the return iron of the superconducting magnet. The
position resolution is about 2 cm.
More details about the design and performance of the detector are given in
Ref.~\cite{Ablikim:2009aa}.

\section{Monte Carlo simulation}
The {\sc geant4}-based~\cite{geant4} Monte Carlo (MC) simulations of $\ee$ annihilations
are used to understand the backgrounds and to estimate detection efficiencies.
The generator {\sc kkmc}~\cite{kkmc} is used to simulate the $\ee$ annihilation incorporating the effects of the beam-energy spread and initial-state radiation (ISR).
The signal modes $\Lambda_c^+\rightarrow \Sigma^{+}\eta^{(\prime)}$ are simulated by taking into account the decay pattern predicted in Ref.~\cite{sharma}, in particular the decay asymmetry parameters are used in the simulation.
The reference modes $\Lambda_c^+\rightarrow \Sigma^{+}\pi^0$ and $\Sigma^{+}\omega$ are simulated according to the 
decay patterns observed in data~\cite{lipr}. 
To study backgrounds, inclusive MC samples consisting of generic $\Lambda_c^+\bar{\Lambda}_c^-$
events, $D_{(s)}^{*}\bar{D}_{(s)}^{(*)}+X$ production, ISR return to the charmonium(-like)
$\psi$ ($Y$) states at lower masses, and continuum processes $\ee\to q\bar{q}~(q=u,d,s)$ are generated, as summarized in Table~\ref{tab:mc_size}.
All decay modes of the $\Lambda_c^+$, $\psi$ and $D_{(s)}$ as
specified in the Particle Data Group (PDG)~\cite{pdg2014} are
simulated with the {\sc evtgen}~\cite{evtgen} generator, while the unknown decays of the $\psi$
states are generated with {\sc lundcharm}~\cite{lundcharm}. 

\begin{center}
\tabcaption{\label{tab:mc_size} Summary of the MC samples size for different processes.}
\footnotesize
\begin{tabular*}{80mm}{l@{\extracolsep{\fill}}c}

\toprule Process        &  Sample size   \\
\hline
	$\Lambda_c^+\rightarrow \Sigma^{+}\eta$ &   0.5M   \\
	$\Lambda_c^+\rightarrow \Sigma^{+}\eta^{(\prime)}$ &   0.5M  \\
 $\Lambda_c^+\bar{\Lambda}_c^-$ (inclusive) &    7.75M \\
	 $D_{(s)}^{*}\bar{D}_{(s)}^{(*)}+X$ &        10.94M\\
   ISR &             4.0M \\
	 $\ee\to q\bar{q}~(q=u,d,s)$ &          277.9M\\

\bottomrule
\end{tabular*}
\vspace{-0mm}
\end{center}

\section{Event selection}

In the selection of $\Lambda_{c}^{+}\rightarrow\Sigma^{+}\eta$, $\Sigma^{+}\eta^{\prime}$, $\Sigma^{+}\pi^0$ and $\Sigma^{+}\omega$ decays, 
the intermediate particles $\Sigma^+$, $\omega$ and $\eta'$ are reconstructed in their decays $\Sigma^{+}\rightarrow p \pi^{0}$, $\omega\rightarrow\pi^+\pi^-\pi^0$ and $\eta'\rightarrow\pi^+\pi^-\eta$,
while the $\eta$ and $\pi^{0}$ mesons are reconstructed in their dominant two-photon decay mode.

For each charged track candidate, the polar angle $\theta$ in the MDC is required to be in the range $|\cos\theta|<0.93$. The distances of closest approach to the interaction point are required to be
less than 10\,cm along the beam direction and less than 1\,cm in the
plane perpendicular to the beam.  The specific ionization energy loss ($dE/dx$) in the MDC and the time of flight information measured in the TOF are
used to calculate particle identification (PID) likelihood values for the pion ($\mathcal{L}_{\pi}$), kaon ($\mathcal{L}_{K}$) and
proton ($\mathcal{L}_p$) hypotheses. Pion candidates are
selected by requiring $\mathcal{L}_{\pi} > \mathcal{L}_{K}$, and proton candidates are required to satisfy $\mathcal{L}_p> \mathcal{L}_{\pi}$ and
$\mathcal{L}_{p} > \mathcal{L}_{K}$.

Photon candidates are reconstructed from isolated clusters in the
EMC in the regions $|\cos\theta| \le 0.80$ (barrel) or $0.86 \le
|\cos\theta|
  \le 0.92$ (end-cap). The deposited energy of a cluster is required to be larger than 25 (50) MeV in the barrel (end-cap) region, and the angle between the photon candidate
and the nearest charged track must be larger than 10$^\circ$. To
suppress electronic noise and energy deposits unrelated to the
events, the difference between the EMC time and the event start time
is required to be within (0, 700)~ns.
Candidates for $\eta$ and $\pi^{0}$ mesons are reconstructed from all $\gamma\gamma$ combinations and the $\gamma\gamma$ invariant mass $M_{\rm \gamma\gamma}$ is required to satisfy $0.50<M_{\rm \gamma\gamma}<0.56\,$GeV/$c^{2}$ for $\eta\rightarrow\gamma\gamma$, and $0.115<M_{\rm \gamma\gamma}<0.150\,$GeV/$c^{2}$ for $\pi^{0}\rightarrow\gamma\gamma$.
 A kinematic fit is
performed to constrain the $\gamma\gamma$ invariant mass to the
nominal mass of $\eta$ or $\pi^0$~\cite{pdg2016}, and the $\chi^2$ of the
kinematic fit is required to be less than 200. The fitted momenta of
the $\eta$ and $\pi^0$ are used in the further analysis.
The invariant masses $M_{p \pi^0}$, $M_{\pi^+\pi^- \pi^0}$ and $M_{\pi^+\pi^- \eta}$ are required to be
within $(1.174,~1.200)$, $(0.760,~0.800)$ and $(0.946,~0.968)$~GeV/$c^2$ for the $\Sigma^{+}$, $\omega$, and $\eta'$ candidates, respectively.

The $\Lambda_c^+$ candidates for all four decay modes are reconstructed by considering all combinations of selected $\Sigma^+$, $\omega$, $\pi^0$ and $\eta^{(\prime)}$ candidates. The $\Lambda^+_c$ candidates are identified based on the beam
constrained mass, $M_{\rm BC} \equiv \sqrt{E^2_{\rm
beam}-|\vec{p}_{\Lambda^+_c}|^2}$, where $E_{\rm
beam}$ is the beam energy and
$\vec{p}_{\Lambda^+_c}$ is the momentum of the
$\Lambda^+_c$ candidate in the rest frame of the initial $e^{+}e^-$ system. To suppress the combinatorial background, a requirement on the
energy difference $\Delta E \equiv E_{\rm beam}-E_{\Lambda^+_c}$ is performed, where
$E_{\Lambda^+_c}$ is the energy of the $\Lambda^+_c$
candidate. 
In practice, to improve the resolution of $\Delta E$, a variable $\Delta Q \equiv \Delta E - k\cdot (M_{p\pi^0} - m_{\Sigma^+})$ is defined  that decouples the correlation between the measured $\Delta E$ and the invariant mass of the $\Sigma^+$ candidate, $M_{p\pi^0}$. Here, $m_{\Sigma^+}$ is the nominal mass of the $\Sigma^+$. 
The factor $k$ is 1.08 for $\Sigma^+\eta$ and $\Sigma^+\pi^0$, and 0.88 for $\Sigma^+\eta^\prime$ and $\Sigma^+\omega$, 
as obtained by a fit to the two-dimensional distributions of  $\Delta E$ versus $M_{p\pi^0}$ with a linear function.

For a specific decay mode, we only keep the candidate with the minimum $|\Delta Q|$ per event. The resultant $\Delta Q$ distribution is shown in Fig.~\ref{fig:deltaq}. 
A mode-dependent $|\Delta Q|$ requirement, which is approximately three times of its resolution and summarized in Table~\ref{tab:sum}, is applied to select candidate signal events.

\end{multicols}

\begin{center}
\begin{minipage}{0.8\textwidth}
\centering
\includegraphics[width=0.46\textwidth]{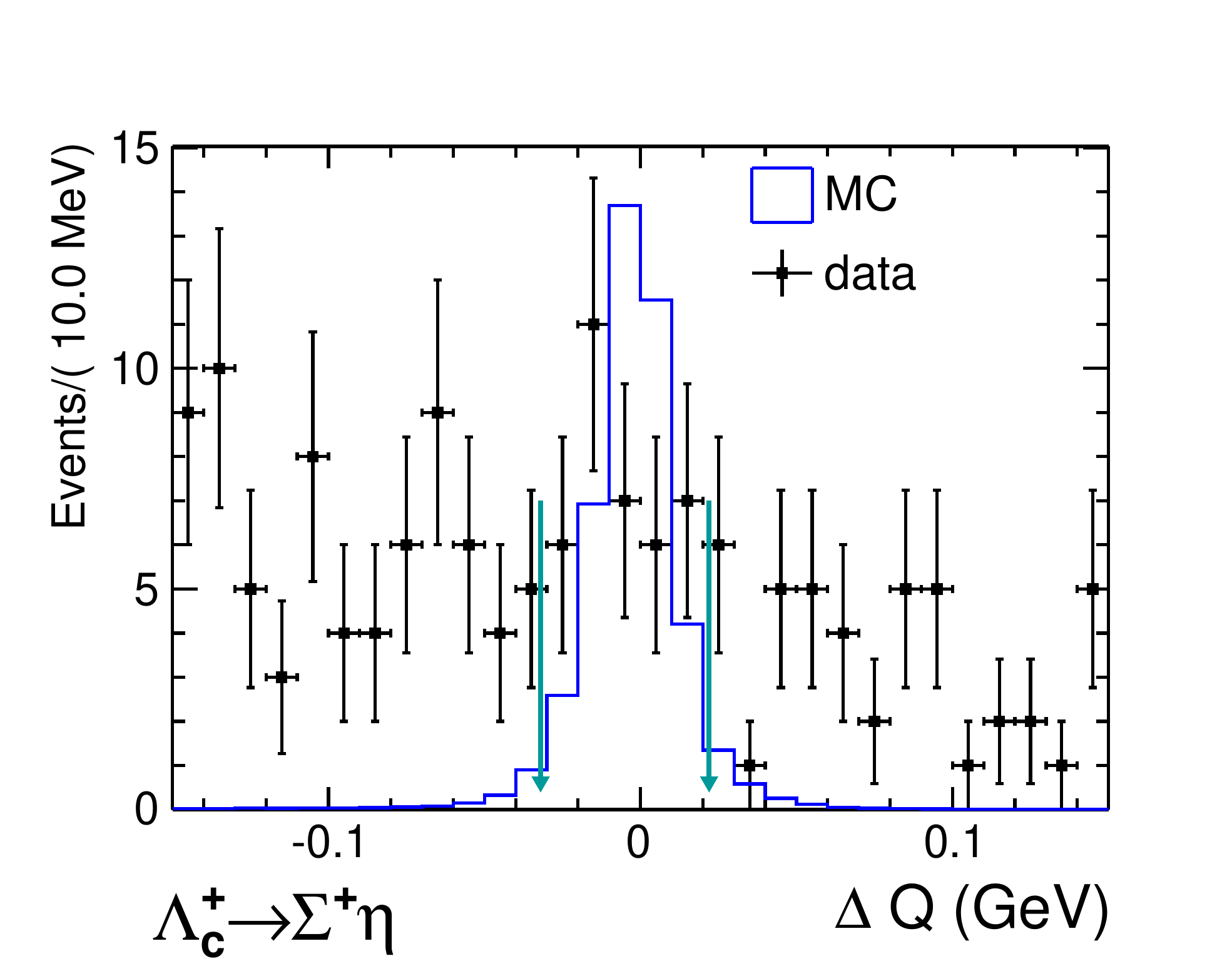}
\put(-135,110){(a)}
\includegraphics[width=0.46\textwidth]{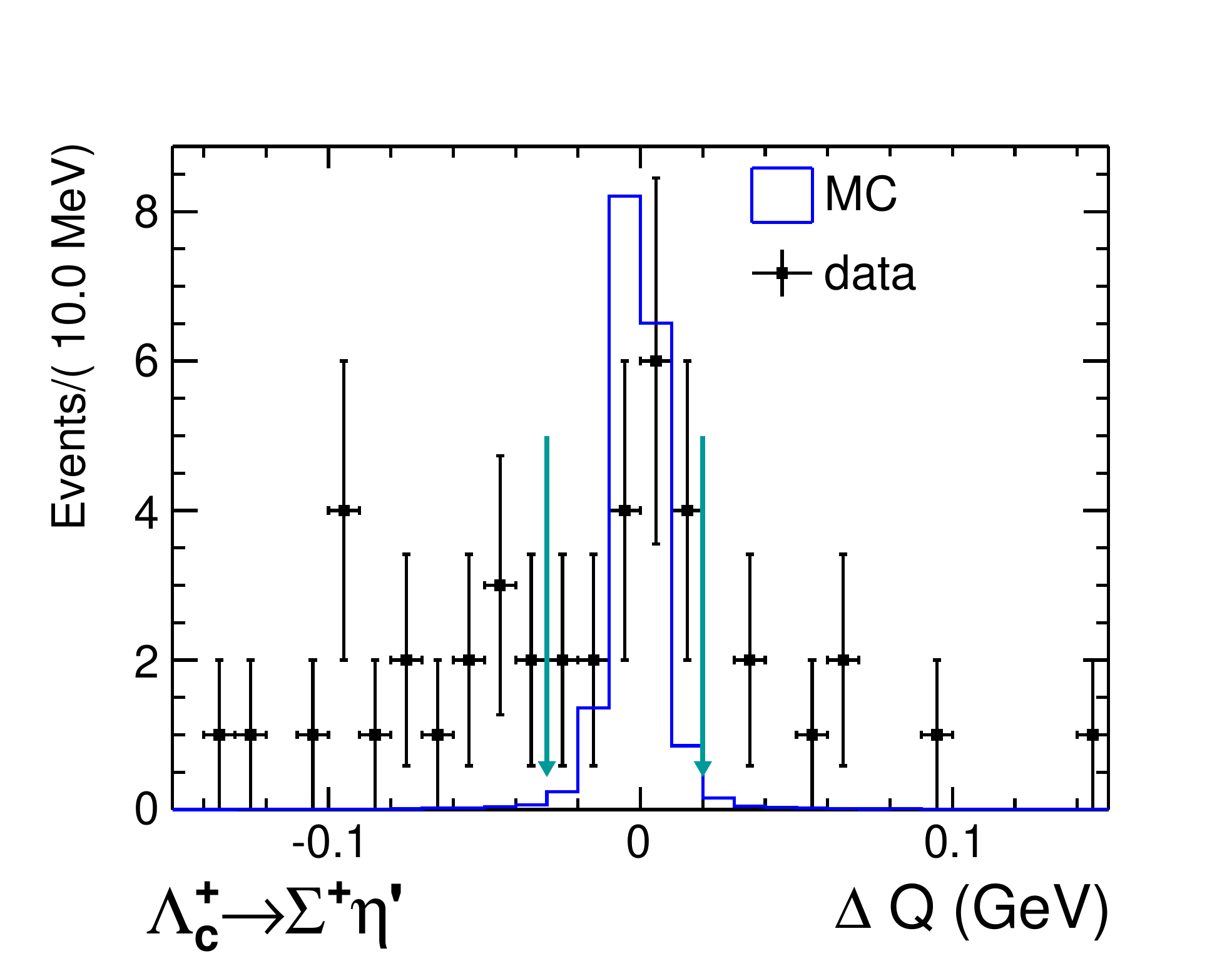}
\put(-135,110){(b)}

\includegraphics[width=0.46\textwidth]{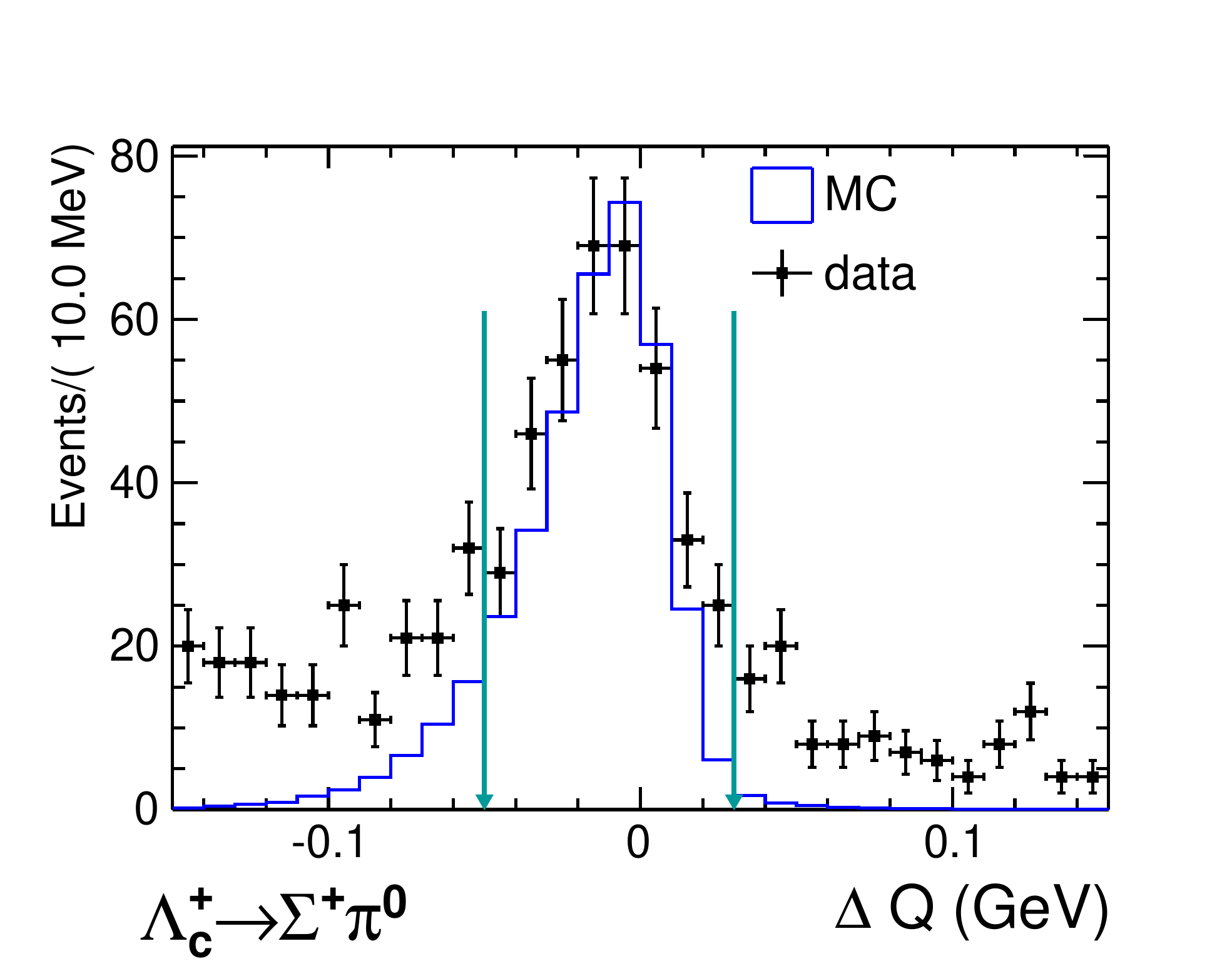}
\put(-135,110){(c)}
\includegraphics[width=0.46\textwidth]{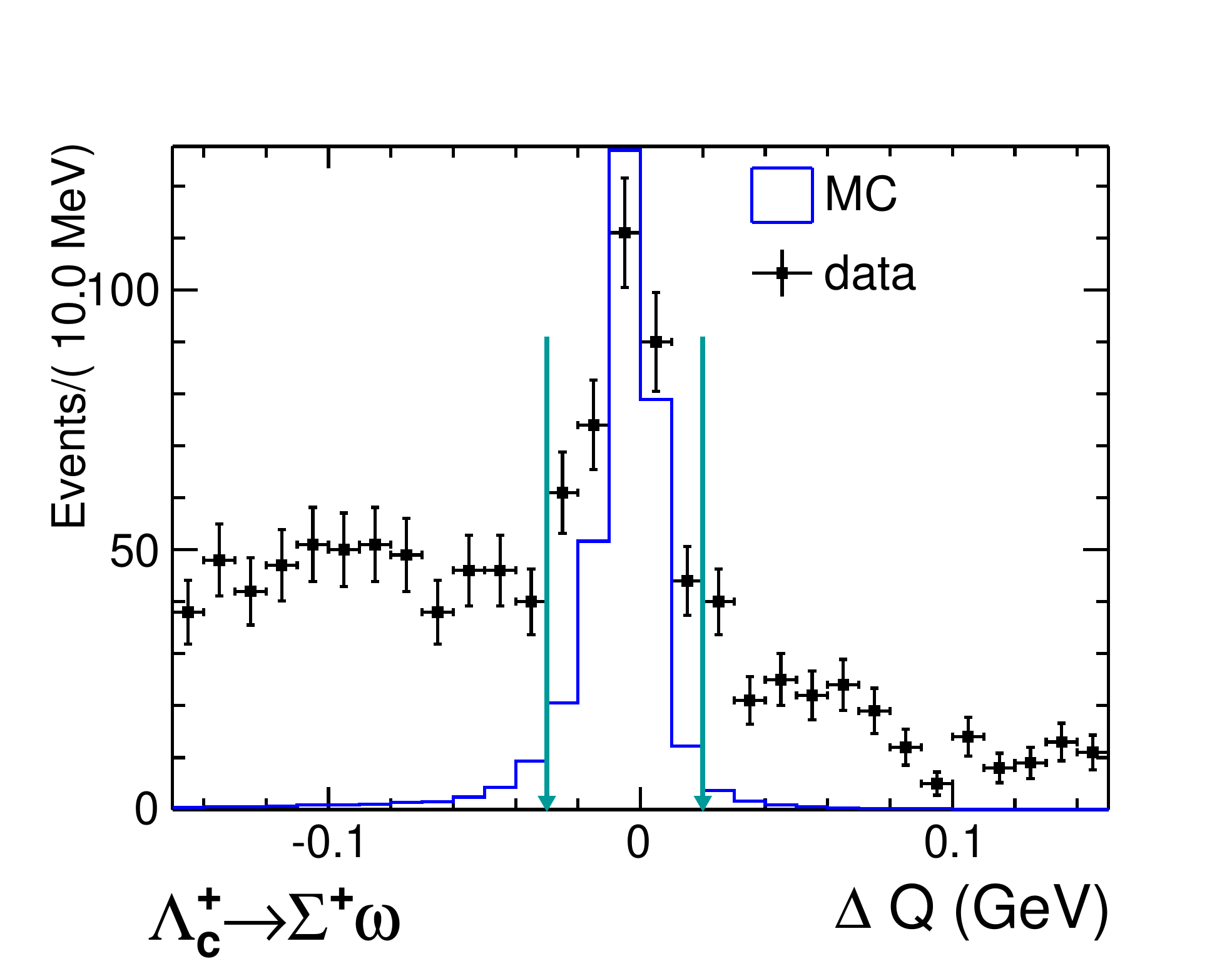}
\put(-135,110){(d)}

\figcaption{\label{fig:deltaq}  Distributions of $\Delta Q$ for $\Lambda_c^+\rightarrow \Sigma^{+}\eta$(a), $\Lambda_c^+\rightarrow \Sigma^{+}\eta'$(b), $\Lambda_c^+\rightarrow \Sigma^{+}\pi^0$(c) and $\Lambda_c^+\rightarrow \Sigma^{+}\omega$(d). Points with error bars are data, solid blue lines are the signal MC samples, the green arrows show the mode-dependent signal region in $\Delta Q$.  The signal MC samples are shown with an arbitrary scale to illustrate the signal shape only.}
 \end{minipage}
 \end{center}

\begin{multicols}{2}

To further suppress the combinatorial backgrounds in the $\Lambda_{c}^{+}\to\Sigma^{+}\eta$ mode, an anti-proton recoiling against the detected $\Lambda_c^+$ candidate is required, which is expected to originate from the $\bar{\Lambda}{}_{c}^{-}$. In order to cancel out systematic uncertainty, the same requirement is applied to the reference mode $\Lambda_{c}^{+}\rightarrow\Sigma^{+}\pi^{0}$.
For the decay mode $\Lambda_c^+\rightarrow\Sigma^+\pi^0$, the peaking background from the CF decay mode $\Lambda_c^+\rightarrow p K_S^0$($K_S^0\rightarrow\pi^0\pi^0$) is rejected by requiring $M_{\pi^0\pi^0}$ not to be in the range (0.48, 0.52) GeV/$c^2$. We also investigate the non-resonant background by checking the $M_{\rm BC}$ distribution of events in the sideband region of the $\Sigma+$, $\eta\prime$ and $\omega$ invariant mass distribution.  No peaking structure from this background is observed.

\end{multicols}

\begin{center}
\begin{minipage}{0.8\textwidth}
\centering
\includegraphics[width=0.46\textwidth]{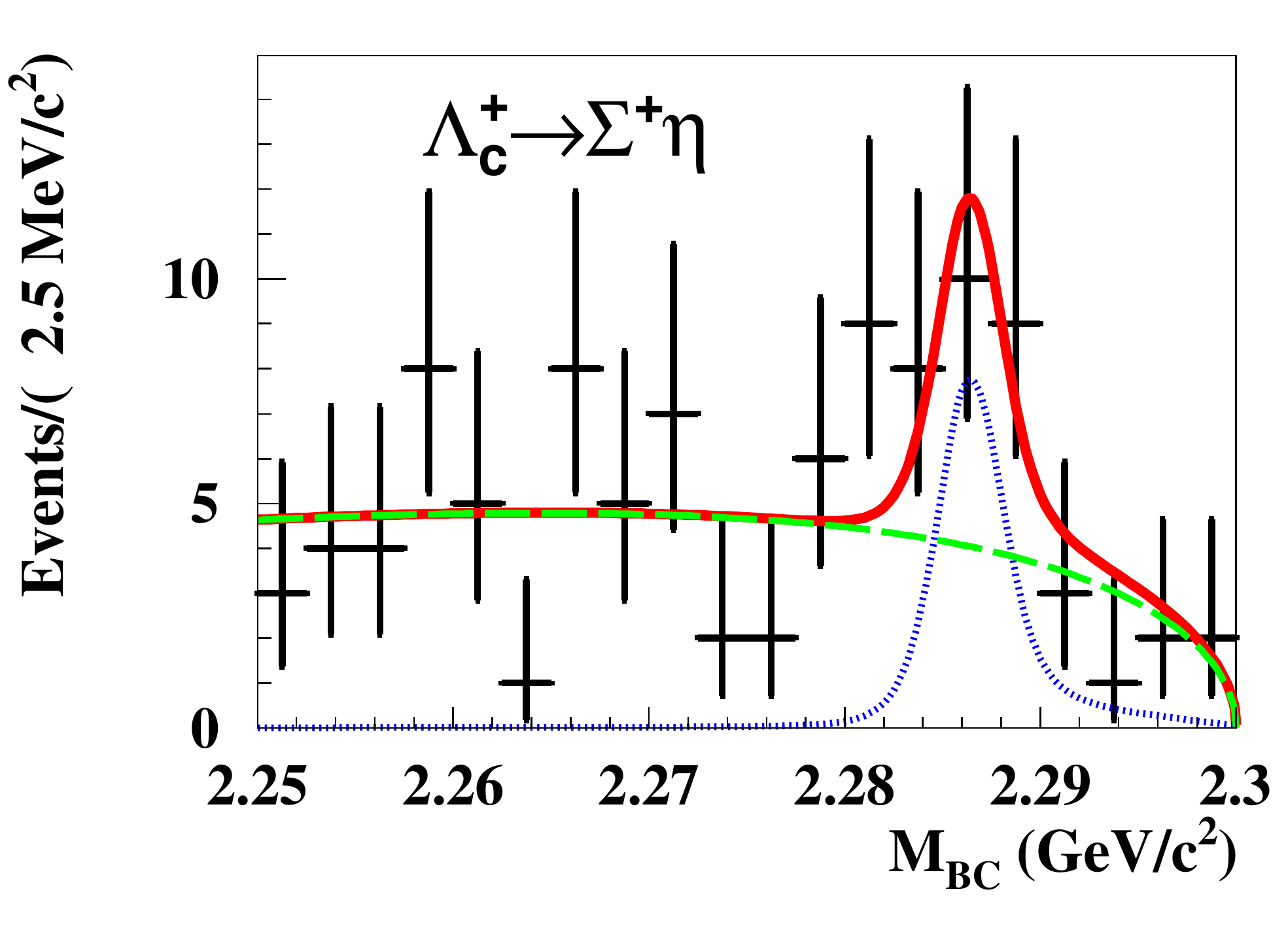}
\put(-135,110){(a)}
\includegraphics[width=0.46\textwidth]{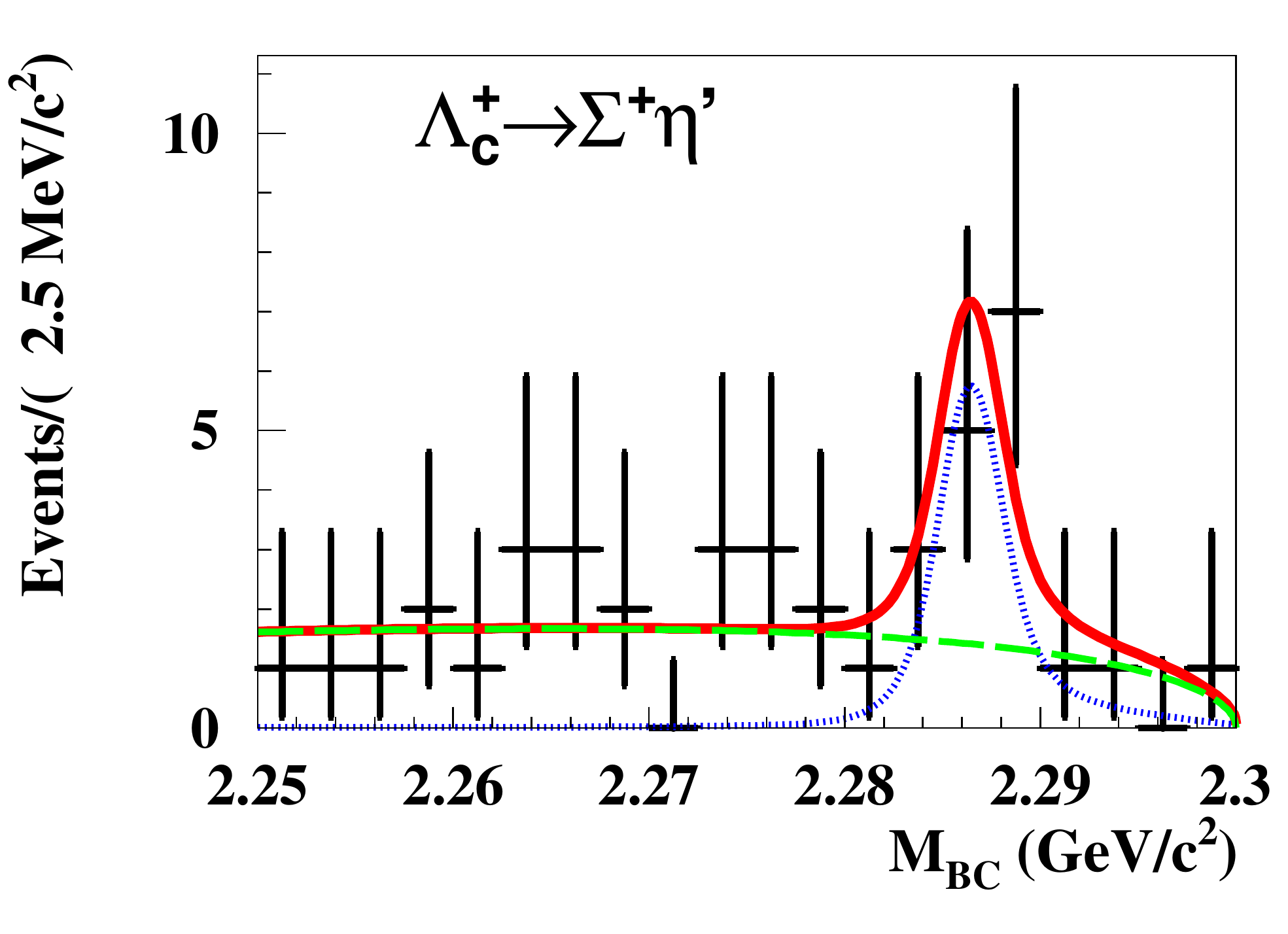}
\put(-135,110){(b)}

\includegraphics[width=0.46\textwidth]{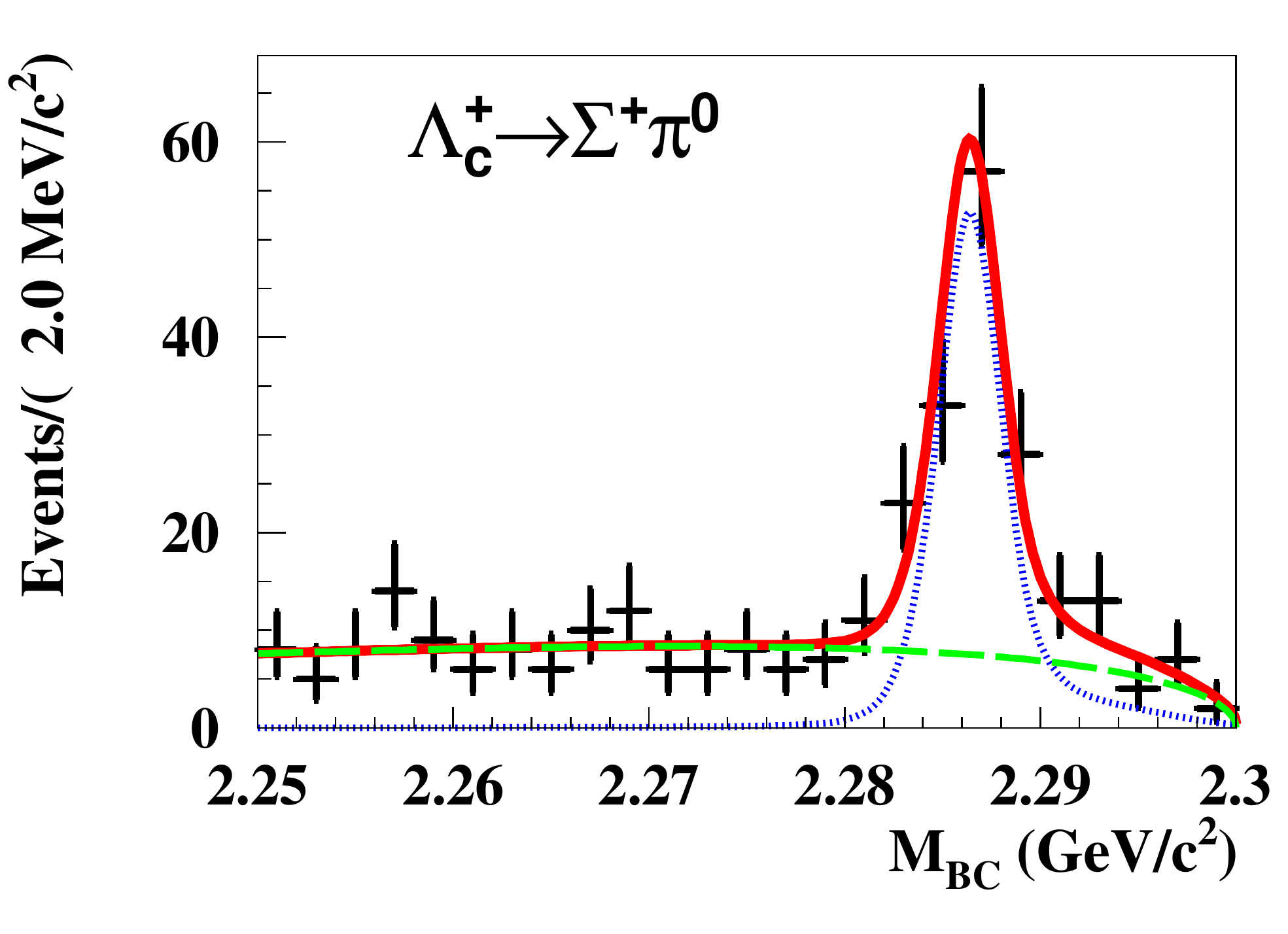}
\put(-135,110){(c)}
\includegraphics[width=0.46\textwidth]{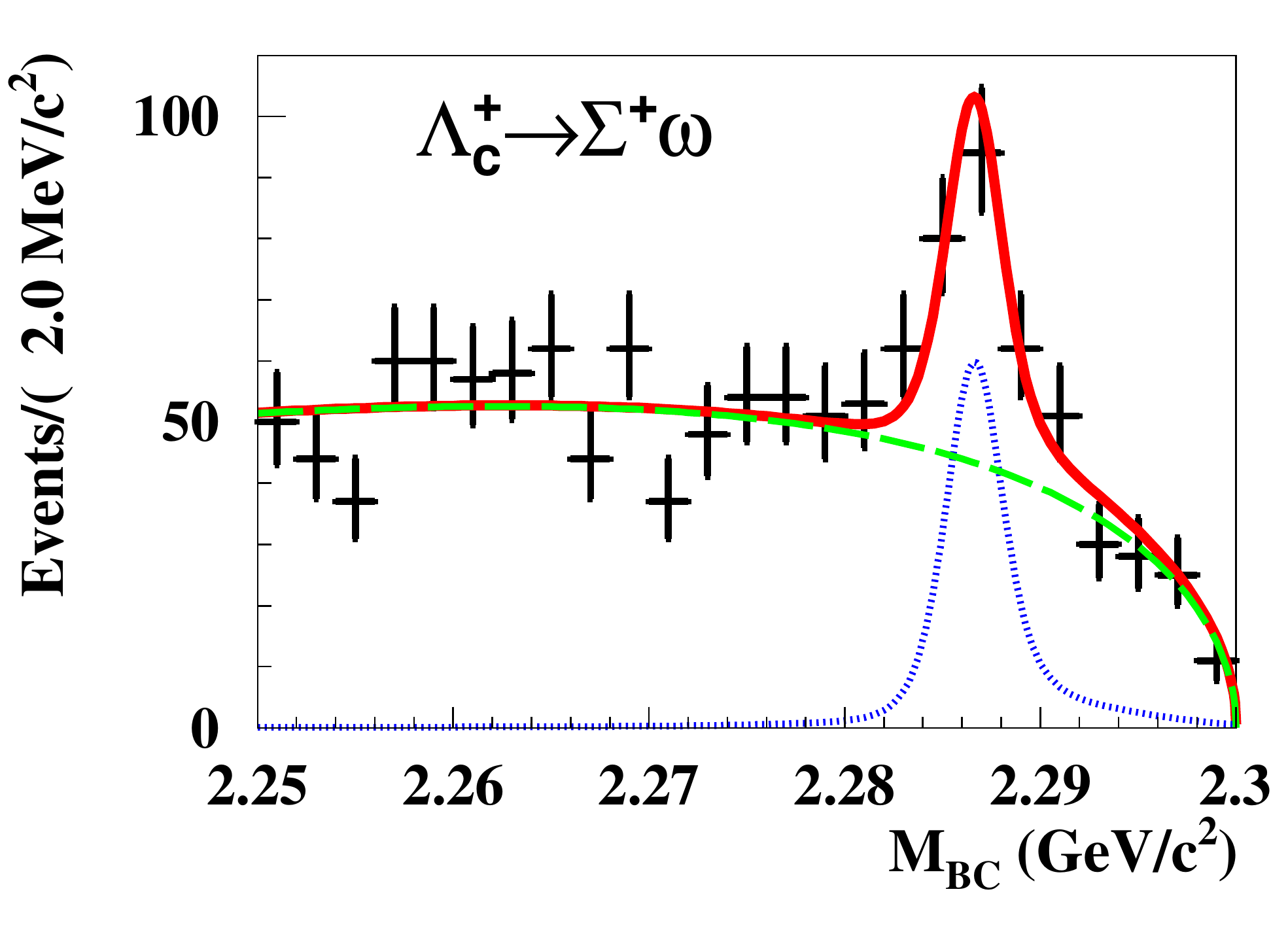}
\put(-135,110){(d)}

\figcaption{\label{fig:total}Fits to the $M_{\rm BC}$ distributions in data for $\Lambda_c^+\rightarrow \Sigma^{+}\eta$(a), $\Lambda_c^+\rightarrow \Sigma^{+}\eta'$(b), $\Lambda_c^+\rightarrow \Sigma^{+}\pi^0$(c) and $\Lambda_c^+\rightarrow \Sigma^{+}\omega$(d). Points with error bars are data, solid lines are the sum of the fit functions,  dotted lines are signal shapes, long dashed lines are the ARGUS functions.}
 \end{minipage}
 \end{center}

\begin{multicols}{2}

\section{Determination of Signal Yields}
After the application of the above selection criteria, 
the $M_{\rm BC}$ distributions of the surviving events are depicted in Figs. \ref{fig:total}(a) and (b) for the signal decay modes $\Lambda_c^+\rightarrow \Sigma^{+}\eta$ and $\Sigma^{+}\eta'$, respectively, and Figs. \ref{fig:total}(c) and (d) for the reference decay modes $\Lambda_c^+\rightarrow \Sigma^{+}\pi^0$ and $\Sigma^{+}\omega$, respectively. 
To determine the signal yields, we perform unbinned maximum likelihood fits to the corresponding $M_{\rm BC}$ distributions. 
In the fit, the signal shapes are described with the MC-simulated signal shapes convolved with a Gaussian function that is used to compensate the resolution difference between data and MC simulations. 
For the signal decay modes, due to the low statistics, the parameters of the Gaussian functions are constrained to those values obtained by fitting the $M_{\rm BC}$ distributions of the corresponding reference decay modes.

The background shapes are modeled with an ARGUS function~\cite{arg}, fixing the high-end cutoff at $E_{\rm beam}$. 
The resulting fit curves are shown in Fig.~\ref{fig:total}, and the signal yields are listed in Table~\ref{tab:sum}. 
The relative ratios of BFs between the signal modes and reference modes are calculated with 

$$R_{ac} = \frac{\br{a}}{\br{c}} = \frac{N_{a}\varepsilon_{c}\br{\pi^0\rightarrow\gamma\gamma}}{N_{c}\varepsilon_{a}\br{\eta\rightarrow\gamma\gamma}} ,      \eqno{(1)} $$
$$R_{bd} = \frac{\br{b}}{\br{d}} = \frac{N_{b}\varepsilon_{d}\br{\omega\rightarrow\pi^{+}\pi^{-}\pi^0}\br{\pi^0\rightarrow\gamma\gamma}}{N_{d}\varepsilon_{b}\br{\eta'\rightarrow\pi^{+}\pi^{-}\eta}\br{\eta\rightarrow\gamma\gamma}} ,      \eqno{(2)}$$
where the indices $a$, $b$, $c$ and $d$ represent the decay modes $\Lambda_c^+\rightarrow \Sigma^{+}\eta$, $\Sigma^{+}\eta'$, $\Sigma^{+}\pi^0$ and $\Sigma^{+}\omega$, respectively. $\br{\pi^0\rightarrow\gamma\gamma}$, $\br{\eta\rightarrow\gamma\gamma}$, $\br{\eta^{\prime}\rightarrow\pi^{+}\pi^{-}\eta}$ and $\br{\omega\rightarrow\pi^{+}\pi^{-}\pi^0}$ are the BFs for $\pi^0$, $\eta$, $\eta^{\prime}$ and $\omega$ decays quoted from PDG~\cite{pdg2016}, $N_{i}$ is the corresponding signal yield and $\varepsilon_i$ is the detection efficiency estimated using MC simulations.
The signal yields and detection efficiencies of the different decay modes are summarized in Table~\ref{tab:sum}.
The resultant ratios are determined to be $R_{ac}=0.35 \pm 0.16$ and $R_{bd}=0.86 \pm 0.34$, where the uncertainties are statistical only.

\begin{center}
\tabcaption{\label{tab:sum} 
Summary of the requirements on $\Delta Q$, signal yields (with statistical uncertainties only) and detection efficiencies for the four decay modes.}

\footnotesize
\begin{tabular*}{80mm}{lccc}

\toprule Decay mode &$\Delta Q$ (GeV) & $N_{i}$ & $\varepsilon_i$ (\%) \\
\hline
    (a) $\Lambda_{c}^{+}\rightarrow\Sigma^{+}\eta$ &[$-$0.032, 0.022] & $14.6 \pm 6.6$ &7.80  \\
	(b) $\Lambda_{c}^{+}\rightarrow\Sigma^{+}\eta'$ &[$-$0.030, 0.020] & $13.0 \pm 4.8$ & 4.61 \\
	(c) $\Lambda_{c}^{+}\rightarrow\Sigma^{+}\pi^{0}$ &[$-$0.050, 0.030] & $122.4 \pm 14.5$ & 8.98\\
	(d) $\Lambda_{c}^{+}\rightarrow\Sigma^{+}\omega$  &[$-$0.030, 0.020] & $135.4 \pm 20.4$ & 7.83\\

\bottomrule
\end{tabular*}
\vspace{0mm}
\end{center}

The statistical significance of the signals for $\Lambda_{c}^{+}\rightarrow\Sigma^{+}\eta$ and $\Sigma^{+}\eta'$ are $2.5\sigma$ and $3.2\sigma$, respectively, which are determined by comparing the likelihood values of the fit with and without the signal component and taking into account the change of the degrees of freedom.

Using the Bayesian method, we set the upper limits at the 90\% confidence level (CL) on the signal yields $N^{\rm UL}_a=24$, corresponding to a ratio of BFs at the 90\% CL $R_{ac}<0.58$ for the decay $\Lambda_{c}^{+}\rightarrow\Sigma^{+}\eta$, and $N^{\rm UL}_b=19$ and $R_{bd}<1.2$ for the decay $\Lambda_{c}^{+}\rightarrow\Sigma^{+}\eta'$.  
The systematic uncertainties discussed below are taken into account by convolving the likelihood curve obtained from the nominal fits with Gaussian functions whose widths represent the systematic uncertainties.

\section{Systematic uncertainty}

Due to the limited statistics, the total uncertainties are dominated by the statistical errors. The systematic uncertainties associated with $\Sigma^+$ detection, tracking and PID of charged pions, and photon selections cancel in the measurement of the ratios of the BFs.

We study the uncertainty associated with the resolution differences between data and MC simulation for $\eta$ and $\pi^{0}$ invariant mass distributions by smearing the $\eta$ and $\pi^{0}$ mass distributions of MC samples with a Gaussian function with a width of 2 MeV/$c^{2}$, as determined by a study of the control channel $D^0 \to K^- \pi^+ \pi^0$. 
The resultant relative changes on the ratios of BFs are 0.3\% for ${R}_{ac}$ and 0.5\% for ${R}_{bd}$ and are taken as the systematic uncertainty due to the different mass resolutions.

We evaluate the uncertainties associated with $\eta'$ and $\omega$ mass requirements with the same method, and the resultant change on ${R}_{bd}$, 0.7\%, is taken as the systematic uncertainty.

The uncertainty related to the $\Delta Q$ requirement is estimated by varying the range by $\pm5$ MeV/$c^{2}$. The corresponding changes, 4.6\% for ${R}_{ac}$ and 6.0\% for ${R}_{bd}$, are taken as the systematic uncertainties.
The uncertainties associated with the fit procedure used to determine the signal yields are studied by performing alternative fits with different fit parameters and fit ranges. More specifically, we vary the values of the two parameters of the Gaussian functions by $\pm1\sigma$, and the fit range by $\pm10$ MeV/$c^{2}$.
Adding the resultant differences in quadrature, we obtain the systematic uncertainty to be 5.9\% and 1.5\% for the ${R}_{ac}$ and ${R}_{bd}$, respectively.

The systematic uncertainties associated with the MC modeling that was used to calculate the detection efficiency are evaluated with different signal MC samples.
 In the nominal analysis, due to limited statistics, the signal MC samples are generated with the helicity angle parameters given in Ref.~\cite{sharma}.  We generate an alternative signal MC sample with additional effects on the decay asymmetry with parameter variations of $\pm$0.2 based on those in Ref.~\cite{sharma}. The resultant changes in the detection efficiencies, which are 2.6\% for ${R}_{ac}$ and 4.4\% for ${R}_{bd}$, are taken as the systematic uncertainties. 

\begin{center}
\tabcaption{\label{tab:sys} Summary of the relative systematic uncertainties in the BF ratio measurements (in unit of \%).}
\footnotesize
\begin{tabular*}{80mm}{l@{\extracolsep{\fill}}cc}
\toprule Source        &  ${R}_{ac}$ & ${R}_{bd}$  \\
\hline
	$\eta'$($\omega$) mass requirement &   -   &   0.7  \\
	$\eta$($\pi^{0}$) mass requirement &   0.3   &   0.5  \\
 $\Delta Q$ requirement &    4.6    &  6.0  \\
	 $M_{\rm BC}$ fit&               5.9    &    1.5  \\

   MC modeling &               2.6 &    4.4    \\
	 MC statistics &           0.2  &         0.2  \\
	 	 ${\mathcal B}_{\rm inter}$ &      0.5   &     1.9    \\
	  \hline
	 Total          &        8.0   &    7.9     \\
\bottomrule
\end{tabular*}
\vspace{-0mm}
\end{center}

The uncertainties of the MC statistics and the decay BFs for the intermediate decays (${\mathcal B}_{\rm inter}$) quoted from the PDG~\cite{pdg2016} are also considered.
All the individual systematic uncertainties are summarized in Table~\ref{tab:sys}. 
The total systematic uncertainties for the measurements of $R_{ac}$ and $R_{bd}$, 8.0\% and 7.9\%, respectively, are obtained by adding the individual values in quadrature.

\begin{table*}[tph!] 
\small
	 \caption{Comparisons of the measured results with theoretical predictions (in unit of \%). \protect\\
	 }
   \begin{center}
   \begin{tabular}{lcccccc}
	 \hline 
	 Decay mode&   K\"{o}rner~\cite{korner} &Sharma~\cite{sharma}  & Zenczykowski~\cite{driver} & Ivanov~\cite{ivan} &  CLEO~\cite{Ammar:1995je} &This work \\  
	 \hline
	 $\Lambda_{c}^{+}\rightarrow\Sigma^{+}\eta$ &  0.16  & 0.57 & 0.94 & 0.11 &0.70$\pm$0.23 & 0.41$\pm$0.20 ($<$0.68) \\
	 $\Lambda_{c}^{+}\rightarrow\Sigma^{+}\eta'$&      1.28 & 0.10 &0.12 &0.12 & -    &  1.34$\pm$0.57 ($<$1.9) \\
	 \hline
	 \end{tabular}
	 
	 \label{tab:vs}
	 \end{center}
	  \end{table*}
\section{Summary}

In summary, by analyzing a data sample of $e^+e^-$collisions corresponding to an integrated luminosity of 567 pb$^{-1}$ taken at a center-of-mass energy of 4.6 GeV with the BESIII detector at the BEPCII collider, we 
find evidence for the decays $\Lambda_c^+\to \Sigma^+\eta$ and $\Sigma^+\eta^\prime$ with statistical significance of 2.5$\sigma$ and 3.3$\sigma$. The BFs for $\Lambda_c^+\to \Sigma^+\eta$ and $\Sigma^+\eta^\prime$ with respect to those of the reference decay modes of $\Lambda_c^+\to \Sigma^+\pi^0$ and $\Sigma^+\omega$ are $\frac{{\mathcal B}(\Lambda_c^+\to\Sigma^+\eta)}{{\mathcal B}(\Lambda_c^+\to\Sigma^+\pi^0)}=0.35\pm0.16\pm0.03$ and $\frac{{\mathcal B}(\Lambda_c^+\to\Sigma^+\eta^\prime)}{{\mathcal B}(\Lambda_c^+\to\Sigma^+\omega)}=0.86\pm0.34\pm0.07$, respectively.
Their 90\% CL upper limits are set to be $\frac{{\mathcal B}(\Lambda_c^+\to\Sigma^+\eta)}{{\mathcal B}(\Lambda_c^+\to\Sigma^+\pi^0)}<0.58$ and $\frac{{\mathcal B}(\Lambda_c^+\to\Sigma^+\eta^\prime)}{{\mathcal B}(\Lambda_c^+\to\Sigma^+\omega)}<1.2$ after taking into account the systematic uncertainties. Incorporating the BESIII results of ${\mathcal B}(\Lambda_c^+\to \Sigma^+\pi^0)$ and ${\mathcal B}(\Lambda_c^+\to \Sigma^+\omega)$ from Ref.~\cite{lipr}, 
we obtain $\br{\Lambda_{c}^{+}\rightarrow\Sigma^{+}\eta}=(0.41\pm0.19\pm0.05)\%$ ($<0.68\%$), and $\br{\Lambda_{c}^{+}\rightarrow\Sigma^{+}\eta'}=(1.34\pm0.53\pm0.21)$\% ($<1.9\%$).

Comparisons of the experimental measurements with theoretical predictions from different models are shown in Table~\ref{tab:vs}. The central value of $\br{\Lambda_{c}^{+}\rightarrow\Sigma^{+}\eta}$ presented in this work is smaller than that from CLEO~\cite{Ammar:1995je}, while they are compatible within $1\sigma$ of uncertainty.
 The BF of $\Lambda_{c}^{+}\rightarrow\Sigma^{+}\eta'$ is measured for the first time, 
which stands a discrepancy about 2$\sigma$ of uncertainty from the most of the theoretical predictions, but in good agreement with the prediction in Ref.~\cite{korner}. 
Furthermore, it is worth noting that the obtained $\br{\Lambda_{c}^{+}\rightarrow\Sigma^{+}\eta'}$ is larger than $\br{\Lambda_{c}^{+}\rightarrow\Sigma^{+}\eta}$, the corresponding ratio is determined to be $\frac{{\mathcal B}(\Lambda_c^+\to\Sigma^+\eta^\prime)}{{\mathcal B}(\Lambda_c^+\to\Sigma^+\eta)}=3.5\pm2.1\pm0.4$, which contradicts with the predictions in Refs.~\cite{sharma, driver}.
However, the precision of the current results is still poor and further constraints demand improved measurements.
\\

\acknowledgments{The BESIII collaboration thanks the staff of BEPCII and the IHEP computing center for their strong support. }

\end{multicols}

\vspace{-1mm}
\centerline{\rule{80mm}{0.1pt}}
\vspace{2mm}

\begin{multicols}{2}

\end{multicols}

\clearpage
\end{CJK*}
\end{document}